\documentstyle[aps,pra,preprint]{revtex}
\begin{document}
%\voffset=-1truein
% \draft command makes pacs numbers print
\draft
\title{Peak effect in the vortex state of V$_3$Si: a study of history dependence}
\author{Sujeet Chaudhary, A. K. Rajarajan, Kanwal Jeet Singh, S. B. Roy
and P. Chaddah}
\address{Low Temperature Physics Laboratory,\\
Centre for Advanced Technology,\\ Indore 452013, India}
\date{\today }
\maketitle
\begin{abstract}
We present results of transport properties measurement 
on a single crystal of V$_3$Si
showing distinct signature
of peak-effect in its vortex state. 
The field variation of the electrical resistance  in the flux-line 
lattice prepared by different experimental path, namely
zero field cooling (ZFC) and field cooling (FC), 
shows a distinct path dependence in the vicinity of the peak-effect regime. 
In the field cooled state, small cycling of magnetic field 
modifies the resistance drastically around the peak-effect
regime, highlighting the metastable nature of that state in the 
concerned regime. 
\end{abstract}                          

\section{Introduction}
The A15 superconductor V$_3$Si has been quite well known 
over the years both for its
interesting normal state \cite{1,2} and superconducting properties \cite{1,3}
and the correlation between the two states \cite{3,4}. There is some
renewed interest in V$_3$Si in recent years, first due to the observation 
of de Haas-van Alphen (dHVA) effect \cite{5,6} , and lately due to the  
suggestion of the magnetic field induced 
phase transition in the flux-lattice structure \cite{7}. 
The observation of  dHVA effect in a superconductor is 
quite puzzling to start with, 
since the superconducting energy gap is likely 
to eliminate quantum oscillations. Recent works on V$_3$Si \cite{5,6,8} 
and various
other superconductors like NbSe$_2$ \cite{9}, Nb$_3$Sn \cite{10}, 
CeRu$_2$ \cite{11},
URu$_2$Si$_2$ \cite{12} have provided more interesting results, but both the 
experimental and theoretical situations are yet to be understood
completely \cite{8}. In a very recent neutron measurement 
it has been observed that in certain field direction,
the hexagonal flux-line lattice (FLL) of V$_3$Si 
distorts with the increase in 
magnetic field and 
abruptly becomes of square symmetry \cite{7}. It is suggested that, 
this transition from the hexagonal to square 
symmetry may be a first order transition \cite{7}. 
Phase transition in flux-line
lattice (FLL) or vortex 
state  in general, has been a subject of much interest in recent years both
from theoretical \cite{13} and  experimental \cite{14} points  of view. In 
clean samples of type-II superconductors with weak pinning 
properties, various
topological phase transitions -- from a quasi-ordered FLL 
(or elastic solid or Bragg-glass) to a flux-line liquid, or
from a quasi-ordered FLL 
to a disordered FLL (or plastic solid or vortex-glass) and then to a flux-line
liquid-- were predicted theoretically 
(see Ref. 15 and references cited therein), 
and have subsequently been 
observed experimentally \cite{16}. 
It is not clear at this moment whether there
is an underlying correlation between the phase transitions associated with the
change in the FLL structure (from hexagonal to square symmetry or
vice versa) and the field-induced change in topological character 
(from ordered/quasi-ordered FLL to disordered FLL and/or flux-line liquid)
of the FLL. For this purpose it is important
to identify various macroscopic as well as microscopic observable associated 
with the proposed phase transitions in the FLL and study those in details.

Peak-effect (PE) is an important observable 
for tracking the  topological phase transitions 
(from ordered FLL to disordered FLL) in various 
high-T$_C$ superconductors (HTSC) \cite{16}. 
PE is actually a generic term used
to describe a peak or local maximum in the field variation of the critical 
current density ( J$_C$(H)) in various type-II superconductors \cite{17}. In
dc magnetization study PE gives rise to a second peak in the field dependence
of magnetization \cite{17}.
PE and its associated features have been used extensively in 
recent years to understand the exact nature of 
FLL phase transitions in various
classes of superconductors including both HTSC \cite{16} as well as 
low T$_C$ \cite{18,19,20,21} materials. 
In a recent theoretical study \cite{22} it is suggested that PE in 
HTSC materials may
be explained by the softening of the FLL due to an underlying structural
phase transition from one FLL symmetry to other. 
This suggestion along with
the experimental observation of the structural transition in the FLL of 
V$_3$Si (Ref.7) have motivated us to study the superconducting mixed state 
properties of V$_3$Si in detail. 
Although there exist
reports of PE in the dc magnetization \cite{23} and transport
measurements \cite{24,25}  of V$_3$Si, 
to our knowledge there
exists no suggestion as yet of any topological phase transition 
(from quasi-ordered elastic FLL to disordered plastic FLL)  
associated with this PE. 

The requirement of a detail study of PE in V$_3$Si has
now become important in the light of various recent developments mentioned 
above. In this paper we shall present results of our transport properties 
measurements  in a good quality single crystal 
sample of V$_3$Si, focussing on PE and various 
interesting features associated with it.
Our results will highlight the field-temperature 
history dependence of PE and associated metastable behaviour in V$_3$Si.
Based on our present results and other relevant experimental information
from the existing literature, 
we shall also discuss the possibility of a phase transition 
in the FLL of V$_3$Si.
           
\section{Experimental}
The V$_3$Si single crystal used in our present study was prepared by Dr. A.
Menovski and it is cut from the same mother ingot, part of which was used 
earlier in de Haas-van Alphen study \cite{6}. While the
residual resistivity ratio 
of the original sample (from which the present sample is cut) was
reported to be 47 (Ref.6), our measurements on the present
sample yield a residual resistivity ratio of 42. 
The electrical transport measurements in
our present study are performed using standard four probe technique. 
We have used  a superconducting magnet and cryostat system (Oxford 
Instruments, UK) to obtain the required 
temperature (T) and field (H) environment. In the configuration of our
measurement the current ( I$_M$ ) is passed along the $<100>$ direction of
the sample and H is applied perpendicular to I$_M$.  The superconducting 
transition temperature (T$_C$) (obtained from our zero field resistance 
measurement) is 16.5K. We have measured the magnetic field dependence of the 
resistance R(H) within the following experimental protocols :
\begin{enumerate}
\item Cool the sample below T$_C$ to various  T of measurement in absence of
any applied field H and then increase H isothermally above the upper critical 
field H$_{C2}$. This is zero-field cool (ZFC) field-ascending mode.
\item After the above step, decrease H isothermally from above H$_{C2}$. This
is ZFC field-descending mode.
\item Field cool the sample in fixed H from a temperature well above T$_C$ to
various temperatures T ($<$T$_C$) of measurement. 
This step is repeated for various H at each T. 
This is field-cool (FC) mode.
\end{enumerate}
\section{Results}
It is well known that for a sample of type-II superconductor with pinning, the
critical current I$_C$ decreases monotonically with the increase in H
and goes to zero at the irreversibility field (H$_{irrv} \leq$H$_{C2}$). 
This is shown schematically in Fig. 1 (a). 
However, for superconductors showing PE,
I$_C$(H) shows a peak or local maximum at an intermediate H value before
finally going to zero at H$_{irrv}$ (see Fig. 1(a)). The signature of PE will
appear in the field dependence of R(H), 
depending on the magnitude of the measuring current I$_M$. If I$_M <$I$_{min}$
(see Fig. 1(a)), R(H) will show zero value up to (depending on the
exact value of I$_M$) almost H$_{Irrv}$
(see Fig. 1(b)). For H$_{irrv}< H <$ H$_{c2}$, R(H) will show flux-flow
resistivity leading to the normal state behaviour for H$>$H$_{C2}$.
On the other hand if I$_M>$I$_{peak}$(H) (see Fig. 1(a)),  
flux-flow resistivity will start at lower field. In either of
these cases, the R(H)
will not bear any signature of the PE. The signature of PE will appear in
R(H) if I$_{min} < I_M < I_{peak}$ (see Fig. 1(a)). In such a situation a 
flux-flow resistance is observed with increasing H where I$_M>$I$_C$(H) but 
R(H) will fall back to zero in the PE regime where I$_M<$I$_C$(H)
before increasing again at higher field (see Fig. 1(b)). 
While a direct measurement of the field dependence of I$_C$
would have been very illuminating for the present study of PE in
V$_3$Si, the lack of a suitable current source (with I$_M >$ 100
mA) constrained us to the study of R(H) only.
Adjusting the measuring current I$_M$ accordingly, 
we present in Fig. 2  
R vs H plots for V$_3$Si at various T showing a 
distinct signature of PE. The (H,T) regime
where PE is observed, roughly agrees with that obtained earlier in magnetic
measurements \cite{23}. The finer quantitative discrepancy 
can be attributed to
the different residual resistivity ratio of the samples used in the earlier
measurement \cite{23} which leads to a small but perceptible 
change in H$_{C2}$(T). (It should be noted here that the normal state
resistance of the present V$_3$Si sample above H$_{C2}$(T)   
is $\approx$ 40 micro ohm as shown in Figs. 2-5.) 

The distinct signature in PE disappears in the H dependence of R(H) 
for T$>$15K. The R vs H plots do not show the {\it reentrant} zero-resistance 
behaviour for any pre-decided 
value of I$_M$ between 10mA and 100mA at T=15.5K, although
a subtle minimum in R(H) is observed for I$_M$=20mA (see Fig.3).It should be
noted here that the existence of PE in V$_3$Si was not at all clear in this
T regime in the magnetic measurements as well \cite{23}.

The PE and the associated features have been very well studied in recent years
in the C15-Laves phase superconductor CeRu$_2$ \cite{18,19,20,26,27}.  
A very distinct
history dependence of PE while cycling H or T 
was observed in CeRu$_2$ and it was suggested 
that at the onset of the PE regime, the  
field cooled (FC) flux-line lattice or the superconducting vortex 
state of CeRu$_2$ was more disordered than the corresponding 
ZFC state \cite{26,27}. We shall now investigate the 
possibility of the same in the
present sample of V$_3$Si. For this, we measure R vs H at various 
T following the
FC protocol (described above in the experimental section), and the results of
such a study are shown in Fig.4. The R(H) measured with the 
FC protocol is found 
to be zero even for H-values at which measuring current 
I$_M$ gave rise to the intermediate H flux-flow regime in the 
measurements following the ZFC protocol (see Fig.2). This clearly indicates 
that in the FC mode I$_C$(H)is greater than 
I$_M$ in this intermediate H regime,  
while  the opposite is true in the ZFC mode. 
Thus I$_C(H)$ is higher in the 
FC mode than in the ZFC mode and this is represented
schematically in Fig.1c. To the best of our knowledge this history dependent 
property of the flux-line lattice of V$_3$Si has not been reported 
in the literature so far.      

The field cooled FLL at the onset of the PE regime in CeRu$_2$ was 
reported to be quite metastable in nature \cite{27}. We shall now focus on the
FC FLL of V$_3$Si and check for the  metastable behaviour in various
H regime. We subject the sample to small field cycling subsequent to the 
initial FC measurement with an applied H at a particular T. The results of
such experiments are shown in Fig.5. We find that within 
the intermediate H regime,
the zero resistance state obtained in the FC mode is destroyed readily on
field cycling and the flux-flow resistance corresponding to ZFC state  is 
recovered. This clearly shows the metastability of 
FC state in the 
intermediate H regime just below the PE regime. We also note 
that the minimum value
of the cycling field ($\Delta H$) 
to destroy the metastable FC state, decreases
as we move away from the PE regime towards lower field. 
While we required $\Delta$H=25 mT at H=3.2T
to reach the flux-flow resistance of the corresponding ZFC state (see Fig.
5), the recorded value of $\Delta H$ at H=2.4T is 5 mT. Below H=1.8T where both
ZFC and FC R(H) show zero resistance, the FC state is quite stable and not
sensitive to any external field cycling. Inside the PE
regime also the field cycling does not have any effect on the FC state. So, 
it is clear that in the intermediate H regime ( where the FC FLL is inferred 
to be more disordered  than the ZFC FLL), the FC state is  metastable
in nature. 
 
\section{Discussion and conclusion}
The main results of our present study are the following:
\begin{enumerate}
\item We see distinct signature of PE in the electrical 
resistance measurements of V$_3$Si.
\item There is a clear path dependence in the electrical resistance in the
field regime at the onset of PE. This path dependence suggests that the FLL 
prepared following the FC protocol is relatively more disordered,
and can carry
more critical current than the FLL prepared in the ZFC protocol.
\item The FC FLL in the intermediate H regime of interest is quite metastable
in nature and is sensitive to external fluctuations in the form of field 
cycling.
\end{enumerate}

A related history dependence of the transport
properties in polycrystalline samples of CeRu$_2$, has earlier been reported
by Dilley et al \cite{28}. These results of transport properties 
measurements are correlated
to the history effects observed in the PE regime in 
various magnetic measurements of CeRu$_2$ \cite{20,26,27}.
We shall now make a comparative study of the history dependence 
of PE in V$_3$Si  with the history effects in 
CeRu$_2$  and NbSe$_2$ where  
it was argued \cite{18,19,20,21,27} that 
the onset of PE marked a field induced  first order transition 
from a relatively ordered FLL to a disordered FLL. While the origin of 
this transition as well as the microscopic nature of the high field-high
temperature  phase
remain a matter of debate (see Ref.20), 
it is generally agreed that on reduction
of H and T the system supercools across 
the first-order transition line \cite{20,21,27,29}.
Experimentally observed field-temperature history effects were associated
with this supercooling effect (see Ref. 29). 
This picture of first order transition is further strengthened with the 
recent theoretical argument (Ref.30) that 
the range of supercooling while reducing T (i.e. in the FC mode) is more than
that obtained through the isothermal variation of H. This is 
observed experimentally both in CeRu$_2$ 
(Ref. 27) and NbSe$_2$ (Ref. 21).
On comparison, the PE and the associated history dependence  in V$_3$Si 
turn out to be very similar to CeRu$_2$ and NbSe$_2$.
However, it should be noted that the history 
dependence of the resistance on isothermal reduction of H (from above 
H$_{C2}$) is relatively 
subtle in V$_3$Si (see inset of Fig.2), 
and  detailed magnetic measurements in the line of those 
in CeRu$_2$ \cite{18,20,27}
are required here. (This will be one of our future projects).

A similar field-temperature history dependence of PE in the 
transport properties
measurements has also been observed in single crystal Nb \cite{31} and thick
film samples of Nb$_3$Ge and Mo$_3$Si \cite{32}. 
In these systems the history dependence
was associated with the pinning properties of the FLL dislocations \cite{32}.
However, while comparing the history effects observed in
CeRu$_2$, we argued (Ref. 20 and 27) that the rather fragile nature of the
history dependent FLL's cannot be explained easily within such a
picture. It would imply that small field excursions anneal
out FLL dislocations or any other source of enhanced pinning.
And this contradicts the conjecture (Ref. 32) that annealing
occurs only when the field is reduced below the peak-effect regime.
The fragile nature of FC FLL is clearly shown in our present 
study on V$_3$Si. We have also shown that the field 
cycling $\Delta$H required to destabilize the 
concerned FC FLL decreases rapidly 
 as one goes away from the PE regime. This is consistent with
the theoretical picture that the energy barrier between the higher 
energy metastable state
and the lower energy stable state in the supercooled regime 
diminishes rapidly on variation of (T,H);
supercooling ceases to exist after (T$^*,H^*$), 
where barrier height goes to zero and the 
metastable state is destroyed with infinitesimal fluctuation \cite{30}.
We believe that the idea of supercooling associated with a
first-order transition probably has an edge over the depinning
or annealing of FLL dislocations in explaining the metastable
behaviour of FC FLL.

Possibility of some kind of a field induced glass-like transition should also 
be considered here, especially when it is known that the low diffusivity of a
glass can cause metastablities. Such metastabilities, however, are associated
with hindered kinetics and not with local minima in free energy. If the 
metastability arises  due to reduced diffusivity, then naive 
arguments suggest that the metastability will be more persistent when larger
motions of particles in configuration space are involved; larger motions are
involved when density is varied , rather than when temperature is varied. In 
the case of FLL a much larger rearrangement of FLL is involved when an (H,T) 
point is reached by varying H isothermally than when the same (H,T) point is
reached by varying T at constant H. Hysteresis would thus be lower in the 
FC case than in the case of isothermal H variation. This is in contrast of
what we have actually observed in the FLL of V$_3$Si in the vicinity 
of PE regime and thus negate the possibility of a glass-like transition to
be associated with the PE in V$_3$Si. 

We would also like to introduce a third possibility, namely an analogy 
(if not exact similarity) with the
random-field Ising systems (RFIM) where similar field-temperature history
effects are well known \cite{34,35}. Most experimental information in RFIM 
systems has been obtained from studies on various diluted antiferromagnets.
In the zero field cooled mode, the diluted antiferromagnet is cooled through
the zero-field Neel transition temperature. The resultant antiferromagnetic 
order thus formed, is preserved when an external magnetic field is 
subsequently switched on at low temperatures. The long range order, however,
gets reduced and ultimately goes to zero on heating the system to the high
temperature paramagnetic phase. On cooling back now from the 
paramagnetic phase in presence of the applied field (i.e. in the field cooled
mode), the sample develops a short range ordered domain state \cite{34} in 
contrast to the long range order ZFC state. The similarity with the FLL
state around the PE regime of V$_3$Si is apparent here, namely the FC FLL 
is relatively more disordered. Like the FC state of RFIMs,
 the FC FLL of V$_3$Si in the vicinity
of PE, also becomes unstable 
on field and temperature cycling \cite{36};
the FC states tend to develop long range order through such cycling
process \cite{36}. It is also interesting 
to note here that the question regarding the
underlying phase transition in RFIMs 
-- whether it is a first order phase transition
or a continuous second order phase transition--is yet to be settled \cite{37}.

We have mentioned in the beginning
that a very recent neutron study has shown the existence of a field induced
structural transition from hexagonal to square symmetry 
in the FLL of V$_3$Si (Ref.7). Such a structural transition is capable of
softening the FLL and can give rise to a PE \cite{22}. 
Moreover, if the structural
transition in the FLL of V$_3$Si is indeed a first order transition as was
speculated in the aforementioned neutron study \cite{7}, then all the 
experimentally observed history dependence of PE in V$_3$Si 
may find a natural explanation in terms of this first order phase 
transition. However, we must note that the reported structural
transition in FLL takes place at a rather low field value of
about 1.3T at 1.8K. Unless there is a strong nonmonotonic
temperature dependence of the onset field of PE in magnetic and
transport measurements as is observed in the case of untwinned
single crystals of YBCO (Ref.38), any simple correlation between PE and
structural transition will not be valid. Future measurements of
magnetic and transport properties down to 1.8K and/or neutrom
measurements in the temperature regime 12K and above will settle
this question.

Softening of the FLL in a type-II superconductor 
may also arise due to a change
in the character of the underlying superconducting order parameter. It was
argued by Fulde and Ferrel \cite{39}, and independently by 
Larkin and Ovchinnikov \cite{40} that in a superconductor with 
substantial normal
state paramagnetism, a partially depaired superconducting state was 
energetically more favorable than the isotropic BCS state. The
superconducting order parameter in this high field state (which is often 
termed as Fulde-Ferrel-Larkin-Ovchinnikov (FFLO) state) vanishes periodically
(with a period of 10-40 times of the coherence length $\xi_C$) along the
direction of the applied field H \cite{41}. These nodes in the order parameter
can transform the three dimensional rigid flux-lines at lower 
H into a quasi-two 
dimensional structure at the onset of the FFLO state\cite{41}
at higher fields. 
A FLL with such a
quasi-two dimensional structure is relatively more soft and can give rise to
a PE. Moreover, since the transition to the FFLO state is a 
first order transition, any system undergoing a FFLO transition should show
along with the PE, the history effects associated with supercooling. 
Although in mid
nineties CeRu$_2$ and the heavy fermion superconductor UPd$_2$Al$_3$ 
were thought to be prime candidates for the FFLO state \cite{42,43}, 
there remain
several arguments  against the existence of such a 
state in those systems (see Ref. 19 and 44). It should be quite 
instructive now to check whether
V$_3$Si meets the following necessary conditions 
for the existence of the FFLO state :
\begin{enumerate}
\item The first condition that such a system 
should be in the clean limit is easily
met for the present single crystal sample of V$_3$Si (Ref.6) as well as in
the other samples of V$_3$Si where PE was observed earlier \cite{23}. 
\item The second
question, whether the upper critical field H$_{C2}$ of V$_3$Si is clearly
Pauli limited or not, is difficult to answer. The 
H$_{C2}(0)\approx$ 18.5 T \cite{6,23} is lower
than the Pauli limiting field H$_P$=1.84 $\times$ T$_C$ Tesla/K $\approx$30T. 
However, the effect of Pauli paramagnetic 
limiting process is thought to be quite
important for the superconducting properties of V$_3$Si \cite{45}. 
It should be
noted here the expression for H$_P$ was derived for a system with spherical
Fermi surface and its application to a system with apparently complicated 
Fermi surface should be made with some caution. It is now known that
the  nesting properties of the Fermi surface actually helps to stabilize 
the FFLO state in a system \cite{46}. Also electron correlation
effect is known to influence the value of H$_P$  in V$_3$Si \cite{47}.

\item While in earlier theoretical works \cite{48} the FFLO state 
ceased to exist for T$\geq$0.57T$_C$, the extent of FFLO state in various
systems is found to be up to $\approx$0.9T$_C$ (Ref. 41).
While PE was not observed in V$_3$Si for T$>$15K 
in magnetic measurements \cite{23},
in our present transport study the field induced zero resistance 
state associated with the PE is observed up to 15K only. However, we have
observed a local minimum (but not the re-entrant zero resistance state) in
the R vs H plot for T$>$15K . This resistance minimum, unlike 
the re-entrant zero resistance in the regime T$\leq$15K, is
path independent.
\end{enumerate}
The above discussion should by no means be taken as a support for the 
existence of the FFLO state in V$_3$Si, but emphasizes that 
more studies, especially 
measurements probing the microscopic properties of the superconducting
states, are required to reach a definite conclusion.

We must mention that a relatively simpler argument for PE is 
also possible in V$_3$Si in terms of minute microscopic 
inhomogeneity present even in the single crystals of V$_3$Si \cite{49}.
Regions with slightly different superconducting properties can act as 
additional pinning centers at high fields, in a similar manner as the 
oxygen deficient centers in the HTSC material YBCO \cite{50}, 
and give rise to PE. However, the magnetic history dependence of PE 
cannot have any simple explanation within such a picture. 

In conclusion, our study of the peak effect in V$_3$Si has revealed 
interesting history effects and metastable features associated with it.
These results can be explained in terms of a first order 
phase transition in the flux-line lattice of V$_3$Si. It is
interesting to note here that very recently history effects and 
associated metastable features have been observed in
the crossover regime from Bragg-glass to vortex-glass in single
crystal samples of YBCO \cite{51}. 
Very recent magneto-optic studies in single crystal samples of
BSCCO claim the presence of phase co-existance \cite{52} and
supercooling across the Bragg-glass to vortex-glass phase
transition \cite{53}. These, in turn, suggest the possibility of a first
order transition. Josephson plasma resonance study in single
crystal samples of BSCCO also support this picture \cite{54}. In
this context, our present results on the well known A15
superconductor V$_3$Si are likely to provide useful information
on the universality of the problem.

\begin{figure}
\caption{(a)Schematic representation of  field ( H ) dependence of 
critical current (I$_C$), 
highlighting the peak-effect near H$_{C2}$. In the 
absence of peak-effect, I$_C$ would have gone to  zero
monotonically (at H$_{irrv}$) as shown by the dashed line. (b)
Schematic representation of field ( H ) dependence of electrical resistance
for various values of the measuring current (I$_M$) -- I$_M<$I$_{min}$, 
I$_{min}<$I$_M<$I$_{peak}$ and I$_M>$I$_{peak}$. 
(c)Schematic representation of 
the field dependence of I$_C$(H) obtained in the ZFC field-ascending mode. 
A history dependence of I$_C$(H) has
actually been observed experimentally in various other type-II
superconductors. For an early detailed study on a strained 
single crystal of Nb 
see M. Steingart, A. G. Putz 
and E. Kramer, J. Appl. Phys. {\bf 44}, 5580 (1973).}
\end{figure} 
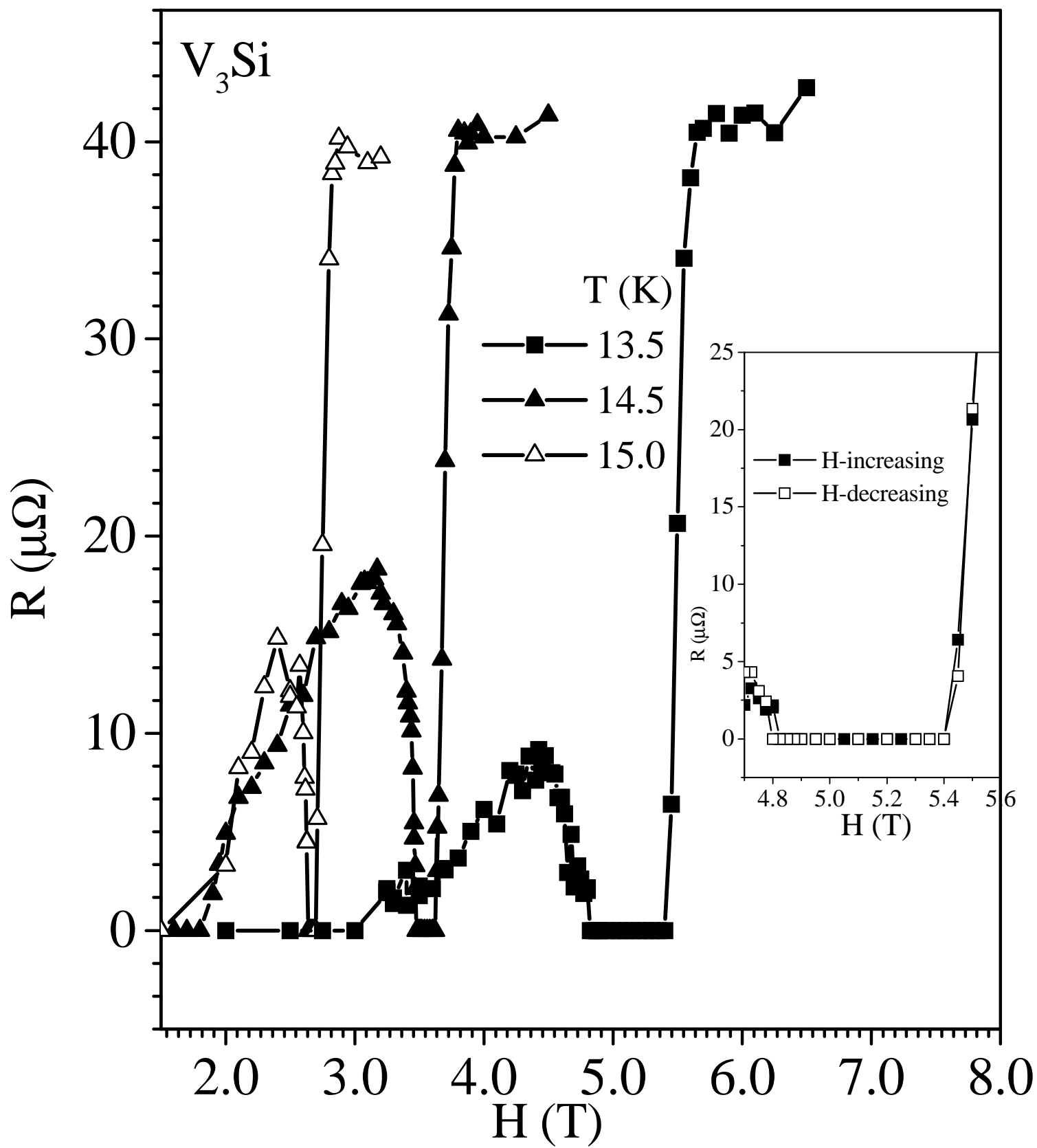
\begin{figure}
\caption{Resistance (R) vs field (H) plot for V$_3$Si 
obtained in the ZFC-field ascending  mode at 13.5K (with  I$_M$ = 100 mA), 
14.5K (with I$_M$=85 mA
 ) and 15 K (with I$_M$=40 mA  ). Inset shows the R vs H plot for V$_3$Si 
obtained in the ZFC-field ascending (filled square)  and ZFC-field 
decscending (open square) mode at 13.5K with  I$_M$ = 100 mA.} 
\end{figure}
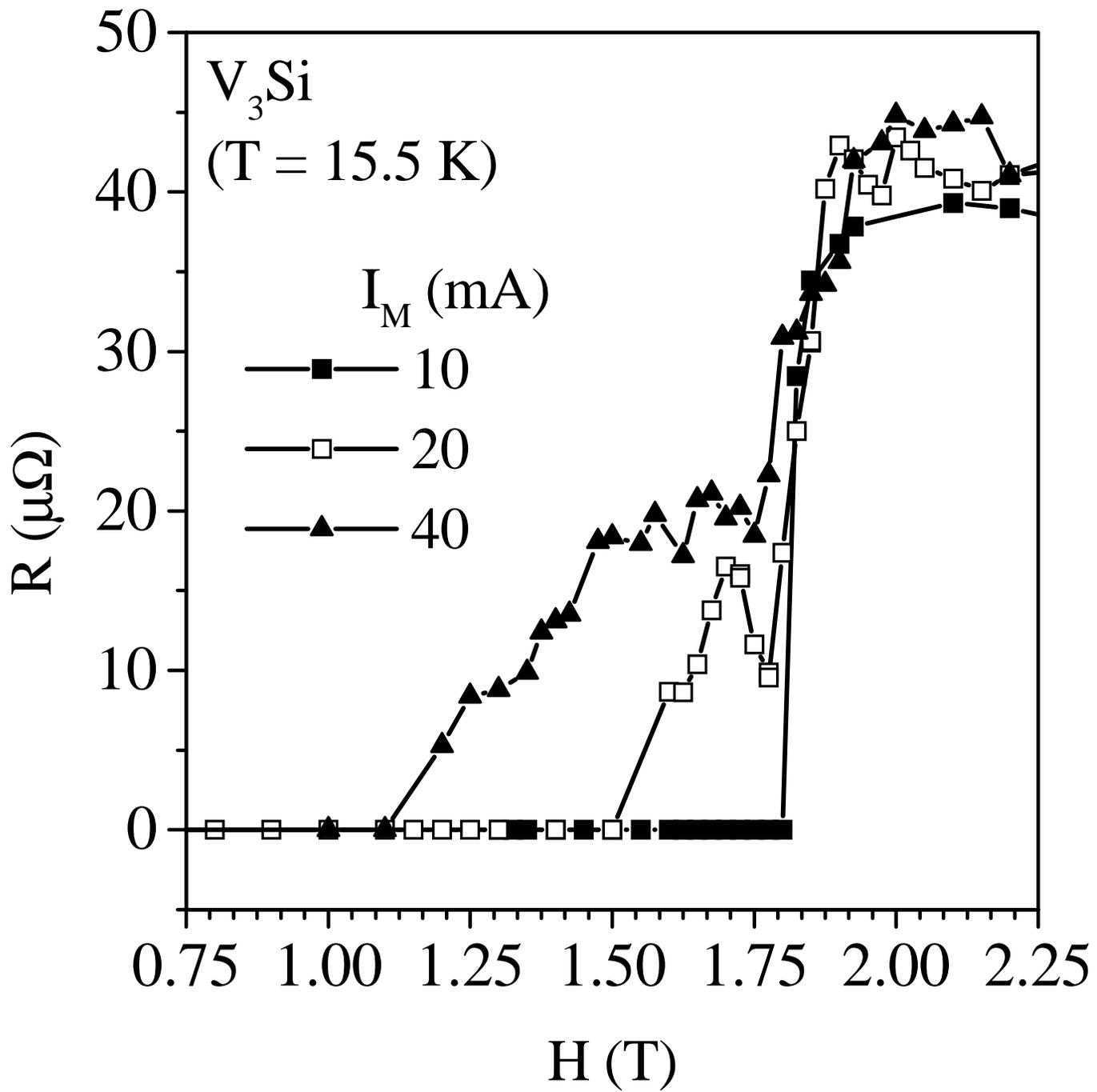
\begin{figure}
\caption{Resistance (R) vs field (H) plot for V$_3$Si 
obtained in the ZFC mode at 15.5K with I$_M$=10, 20 and 40 mA.}
\end{figure}
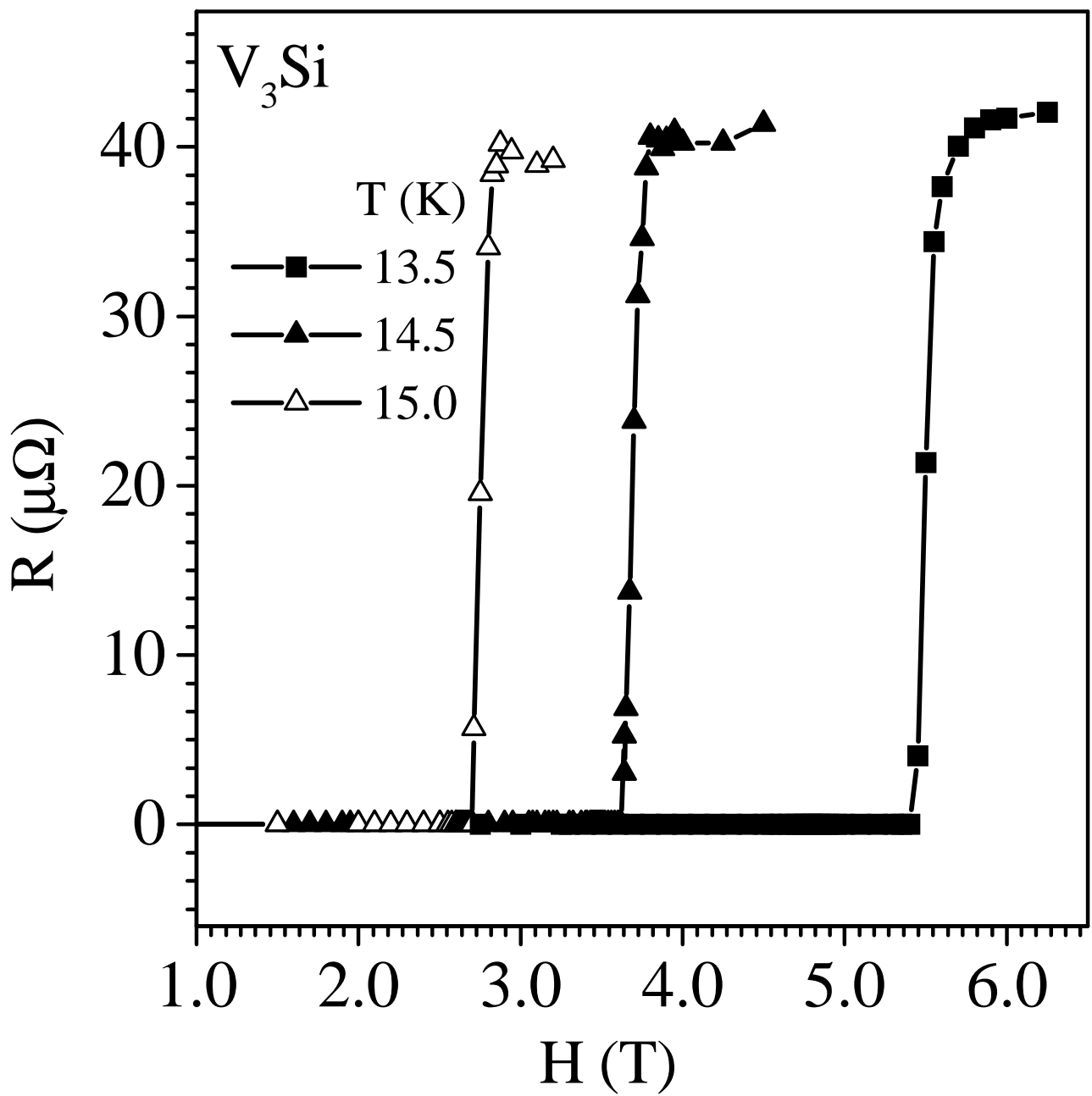
\begin{figure}
\caption{Resistance (R) vs field (H) plot for V$_3$Si 
obtained in the FC mode at 13.5K (with  I$_M$ = 100 mA), 14.5 K (with I$_M$=85 mA
 ) and 15 K (with I$_M$=40 mA).}
\end{figure}
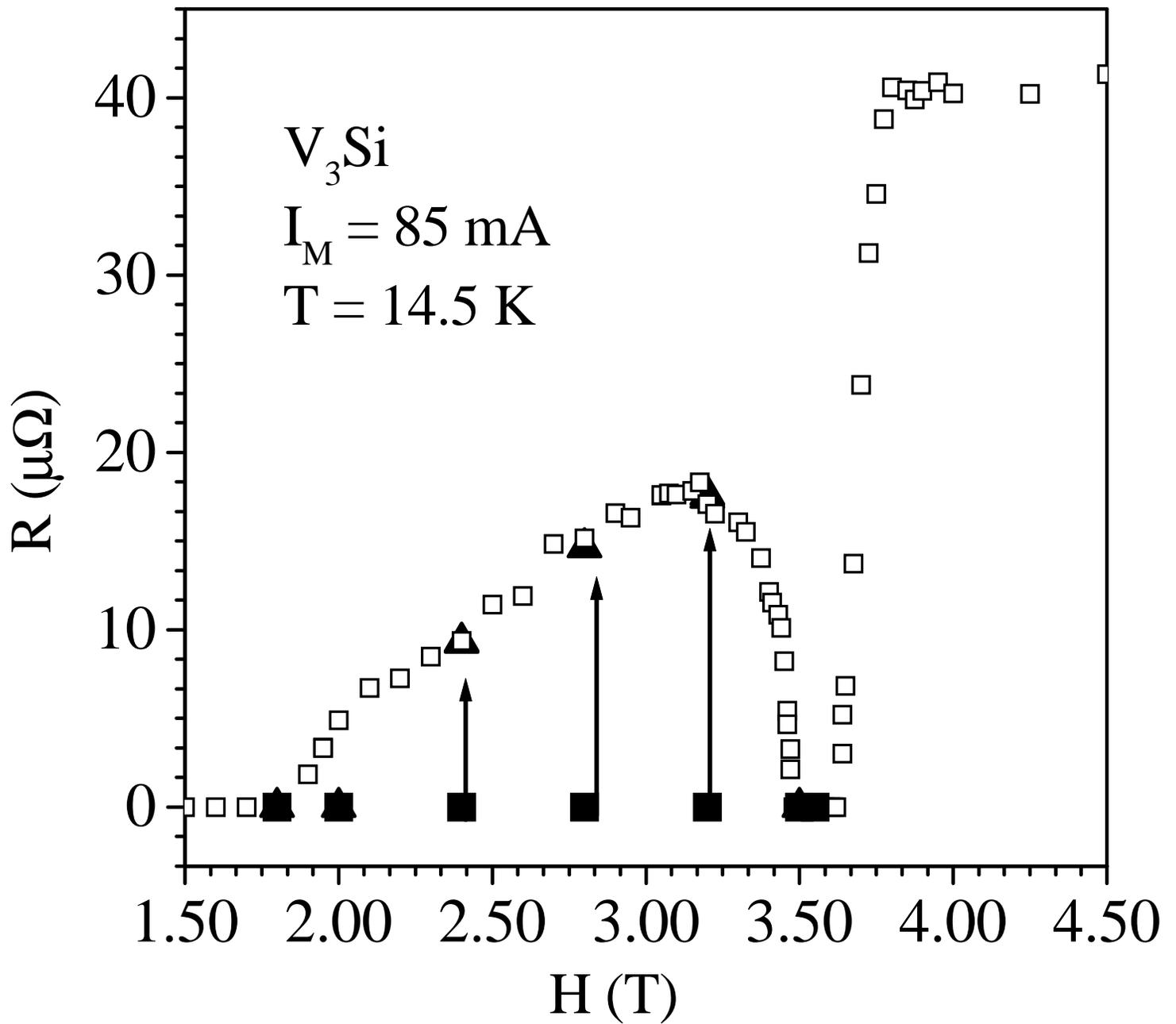
\begin{figure}
\caption{Metastable behaviour of R(H) of 
V$_3$Si obtained in the FC mode
at 14.5K with I$_M$ =85 mA. Filled square denote R(H) values obtained after 
field cooling in various H values. Filled triangles denote R(H) values obtained 
after a field cycling of maximum $\Delta$H 
subsequent to the first FC measurement
at the corresponding H. The magnitude of $\Delta$H depends on H (see text 
for details). Open square denotes data obtained in the ZFC mode at 14.5K ( as in the case of Fig.2). }
\end{figure}

\begin{references}
\bibitem{1}L. R. Testardi, Rev. Mod. Phys. {\bf 47}, 637 (1975).
\bibitem{2}P. W. Anderson, A Career in Theoretical Physics
(World Scientific, 1994) p464.
\bibitem{3}S. V. Vonsovsky, Yu. A. Izyumov and E. Z. Kurmaev, 
Superconductivity of transition metals (Springer Verlag, 1982).
\bibitem{4}G. Bilbro and W. L. McMillan, Phys. Rev. {\bf B14}, 1887 (1976).
\bibitem{5}F. M. Mueller, D. H. Lowndes, Y. Y. Chang, A. J.
Arko and R. S. List, Phys. Rev. Lett. {\bf 68}, 3928 (1992).
\bibitem{6}R. Corcoran, N. Harrison, S. M. Hayden, P. Meeson, M.
Springford and P. J. van der Wel, Phys. Rev. Lett. {\bf 72}, 701 (1994).
\bibitem{7}M. Yethiraj, D. K. Christen, D. McK. Paul, P. Miranovic and J. R.
Thompson, Phys. Rev. Lett. {\bf 82}, 5112 (1999).
\bibitem{8}T. J. B. M. Janssen, C. Haworth, S. M. Hayden, P.
Meeson, M. Springford, A. Wasserman, Phys. Rev. {\bf B57} 11698 (1998).
\bibitem{9}R. Corcoran, P. Meeson, Y. Onuki, P. A. Probst, M.
Springford, K. Takita, H. Harima, G. Y. Guo and B. L. Gyorffy,
 J. Phys.:Condens Matter {\bf 6} 4479 (1994).
\bibitem{10}N. Harrison, S. M. Hayden, P. Meeson, M. Springford,
P. J. van der Wel and A. Menovsky, Phys. Rev. {\bf B50}, 4208 (1994).
\bibitem{11}M. Hedo, Y. Inada, T. Ishida, E. Yamamoto, Y. Haga,
Y. Onuki, M. Higuchi and A. Hasegawa, J. Phys. Soc. Jpn. {\bf 64}, 4535 (1995).
\bibitem{12}H. Ohkuni, T. Ishida, Y. Inada, Y. Haga, E.
Yamamoto, Y. Onuki and S. Takahashi, J. Phys. Soc. Jpn. {\bf 66} 945 (1997).
\bibitem{13}G. Blatter, M. V. Feigekman, V. B. Geshkenbein, A.
I. Larkin, and V. M. Vinokur, Rev. Mod. Phys. {\bf 66} 1125 (1994)
\bibitem{14}E. Zeldov, D. Majer, M. Konczykowski, V. B.
Geshkenbein, V. M. Vinokur and H. Shtrikman, Nature {\bf 375} 373 (1995)
\bibitem{15}T. Giamarchi and P. Le Doussal, Phys. Rev. {\bf B52}, 1242 (1995)
\bibitem{16}B. Khaykovich, E. Zeldov, D. Majer, T. W. Li, P. H.
Kes, and M. Konczykowski, Phys. Rev. Lett. {\bf 76}, 2555 (1996); 
K. Deligiannis, P. A. J. de Groot, M. Oussena, S. Pinfold, R.
Langan, R. Gagnon and L. Taillefer, Phys. Rev. Lett. {\bf 79}
2121 (1997); D. Giller, A. Shaulov, R. Prozorov, Y. Abulafia, Y.
Wolfus, L. Burlachkov, Y. Yeshurun, E. Zeldov, V. M. Vinokur, J.
L. Peng and R. L. Greene, Phys. Rev. Lett. {\bf 79} 2542 (1997).
\bibitem{17}A. M. Campbell and J. E. Evetts, Adv. Phys. {\bf 21} 327 (1972).
\bibitem{18}S. B. Roy and P. Chaddah, Physica {\bf C279} 70 (1997).
\bibitem{19}S. B. Roy and P. Chaddah, J. Phys.:Condensed Matter {\bf 9} (1997)
L625.
\bibitem{20}S. B. Roy, P. Chaddah and S. Chaudhary, J. Phys.:Condensed Matter,
{\bf 10} 4885 (1998).
\bibitem{21}G. Ravikumar, P. K. Mishra, V. C. Sahni, S. S.
Banerjee, S. Ramakrishnan, A. K. Grover, P. L. Gammel, D. J.
Bishop,  E. Bucher, M. J. Higgins and S. Bhattacharya, Physica C
 {\bf 322} 145 1999.
\bibitem{22}B. Rosenstein and A. Knigavko, Phys. Rev. Lett. 
{\bf 83}, 844 (1999)
\bibitem{23}M. Isino, T. Kobayashi, N. Toyota, T. Fukase and Y.
Muto, Phys. Rev. {\bf B38} 4457 (1988).
\bibitem{24}M.Pulver, Z. Physik {\bf 257} 22 (1972).
\bibitem{25}R. Meier-Hirmer, H. Kupfer and H. Scheurer, Phys.
Rev. {\bf B31} 183 (1985). 
\bibitem{26}G. Ravikumar, V. C. Sahni, P. K. Mishra, T. V. C.
Rao, S. S. Banerjee, A. K. Grover, S. Ramakrishnan, S.
Bhattacharya, M. J. Higgins, E. Yamamoto, Y. Haga, M. Hedo, Y.
Inada and Y. Onuki, Phys. Rev. {\bf B57} R11069 (1998).
\bibitem{27}S. B. Roy, P. Chaddah and S. Chaudhary, J. Phys.:Condensed Matter,
{\bf 10} 8327 (1998).
\bibitem{28}N. R. Dilley, J. Hermann, S. H. Han and M. B. Maple,
 Phys. Rev. {\bf B56} 2379 (1997).
\bibitem{29}P. Chaddah and S. B. Roy, Bull. Matr. Sci. {\bf 22} 275 (1999).
\bibitem{30}P. Chaddah and S. B. Roy, Phys. Rev. {\bf B60} 11926 (1999) .
\bibitem{31}M. Steingart, A. G. Putz and E. J. Kramer, 
J. Appl. Phys. {\bf 44} 5580 (1973).
\bibitem{32}R. Wordenweber, P. H. Kes and C. C. Tsuei, 
Phys. Rev. {\bf B33} 3172 (1986).
\bibitem{33}D. P. Belanger, Phase Transitions {\bf 11} 53 (1988).
\bibitem{34}J. A. Mydosh, Spin-glasses (Taylor and Francis, 1993).
\bibitem{35}R. J. Birgeneau, J. Magn. Magn. Mater. {\bf 177} 1 (1998).
\bibitem{36}S. Chaudhary, A. K. Rajarajan, K. J. Singh, S. B. Roy and 
P. Chaddah, Solid St. Commun. (in press).
\bibitem{37}J. P. Hill, Q. Feng, Q. J. Harris, R. J. Birgeneau,
A. P. Ramirez and A. Cassanho, Phys. Rev. {\bf B55} 356 (1997);
 Q. Feng, Q. J.Harris, R. J. Birgeneau and J. P. Hil,
Phys. Rev. {\bf 55} 370 (1997). 
\bibitem{38}D. Giller, A. Shaulov, Y. Yeshurun, J. Giapintzakis,
 Phys. Rev. {\bf B60} 106 (1999).
\bibitem{39}P. Fulde and R. A. Ferrel, Phys. Rev. {\bf 135} A550 (1964).
\bibitem{40}A. I. Larkin and Y. N. Ovchinnikov, Sov. Phys. JETP {\bf 20} 7
62 (1965).
\bibitem{41}M. Tachiki, S. Takahashi, P. Gegenwart, M. Weiden,
C. Geibel, F. Steglich, R. Modler, C. Paulsen and Y. Onuki,
 Z. Phys. {\bf B100} 369 (1996).
\bibitem{42}R. Modler, P. Gegenwart, M. Lang, M. Deppe, M.
Weiden, T. Luhmann, C. Geibel, F. Steglich, C. Paulsen, J. L.
Tholence, N. Sato, T. Komatsubara, Y. Onuki, M. Tachiki, and S.
Takahashi, Phys. Rev. Lett. {\bf 76} 1292 (1996).
\bibitem{43}F. Steglich, R. Modler, P. Gegenwart, M. deppe, M.
Weiden, M. Lang, C. Geibel, T. Luhmann, C. Paulsen, J. L.
Tholence, Y. Onuki, M. Tachiki and S. Takahashi, Physica
 {\bf C263} 498 (1996).
\bibitem{44}N. Dilley and M. B. Maple, Physica {\bf C278} 207 (1997).
\bibitem{45}T. P. Orlando, E. J. Mc Niff Jr., S. foner and M. R.
Beasley, Phys. Rev. {\bf B19} 4545 (1979)
\bibitem{46}H. Simahara, Phys. Rev. {\bf B50} 12760 (1994).
\bibitem{47}T. P. Orlando and M. R. Beasley, Phys. Rev. Lett. {\bf 46} 1598
(1981)
\bibitem{48}L. W. Gruenberg and L. Gunther, Phys. Rev. Lett. {\bf 16} 996 (1966)
\bibitem{49}P. Chaddah and R. O. Simmons, Phys. Rev. {\bf B27} 119 (1983).
\bibitem{50}M. Daumling, J. M. Seuntjens and D. C.Larbelstier, 
Nature, {\bf 346} 332 (1990).
\bibitem{51} S. Kokkaliaris, P. A. J. de Groot, S. N. Gordeev, A. A.
Zhukov, R. Gagnon, L. Taillefer, Phys. Rev. Lett. {\bf 82} 5116
(1999); 
S. Kokaliaris, A. A. Zhukov, P. A. J. de Groot, R.
Gagnon, L. Taillefer and T. Wolf, Phys. Rev. {\bf B61} 3655
(2000); S. O. Valenzuela and V. Bekeris, Phys. Rev. Lett, {\bf
84} 4200 (2000); Y. Redzyner, S. B. Roy, D. Giller, Y. Wolfus, A.
Shaulov, P. Chaddah and Y. Yeshurun, Phys. Rev. {\bf B} to
appear in 1 June 2000 issue.
\bibitem{52}D. Giller, A. Shaulov, T. Tamegai and Y. Yeshurun,
 Phys. Rev. Lett. {\bf 84} 3698 (2000).
\bibitem{53}C. J. van der Beek, S. Colson, M. V. Indenbohm and M.
Konczykowski, Phys. Rev. Lett. {\bf 84} 4196 (2000).
\bibitem{54}M. B. Gaifullin, Y. Matsuda, N. Chikumoto, J. Shimoyama and K.
Kishio, Phys. Rev. Lett. {\bf 84} 2945 (2000).

\end{references}
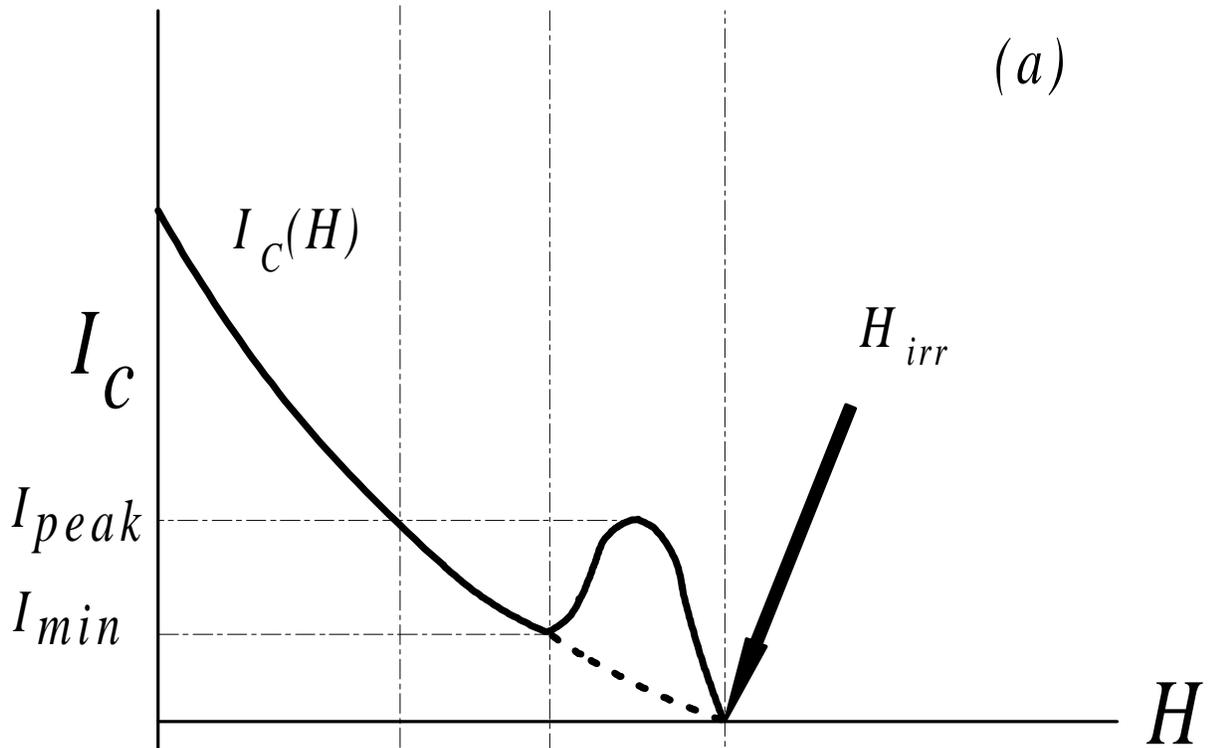
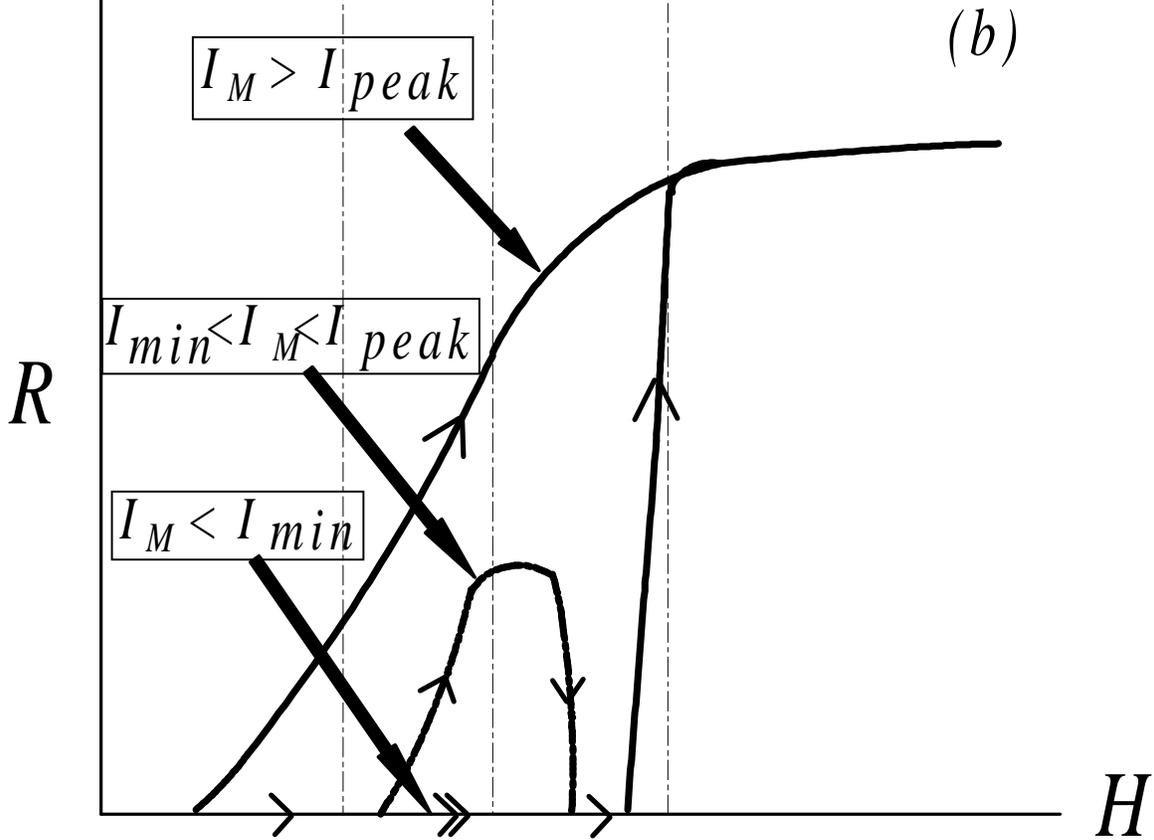
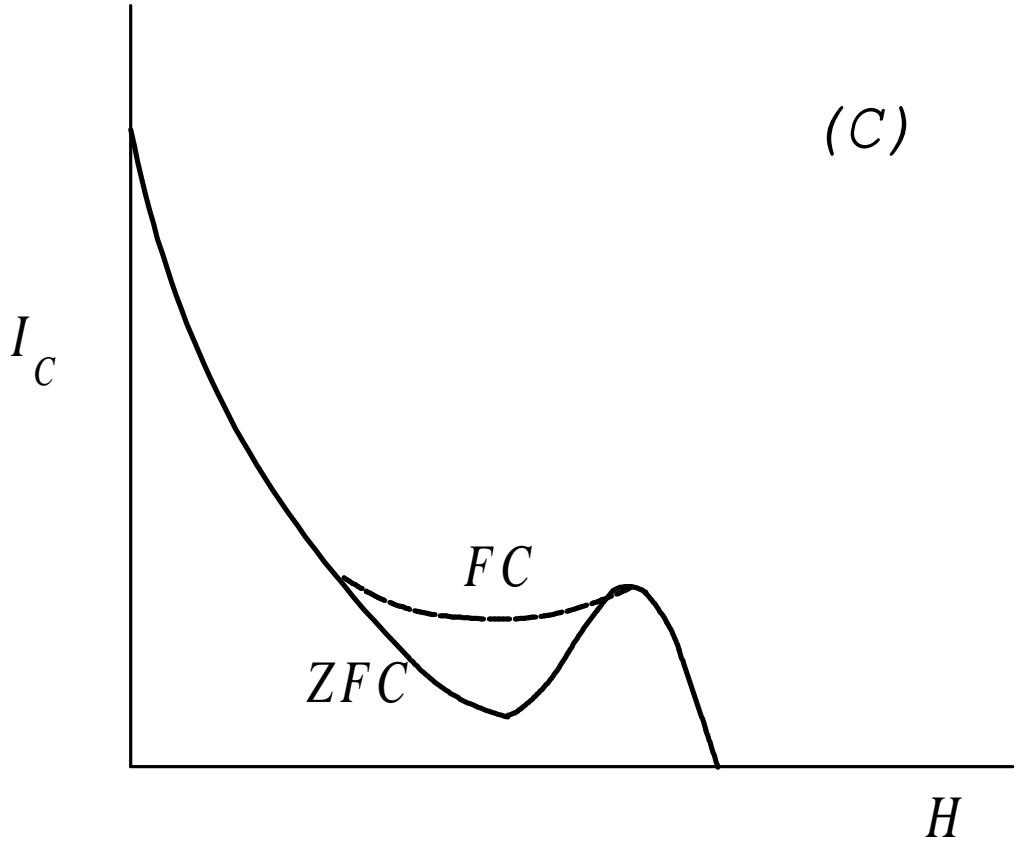
\end{document}